\documentclass[12pt]{iopart}

\usepackage{graphicx}

\begin{document}

\title[PHENIX Jets and $\gamma$-Hadrons]{Probing Nuclear Matter With Jets and $\gamma$-Hadron Correlations: Results from PHENIX}
\author{N.~Grau (for the PHENIX Collaboration\footnote{A list of members of the PHENIX
Collaboration can be found at the end of this issue.})}
\address{Columbia University, Nevis Laboratories, Irvington, NY, 10533, USA}
\ead{ncgrau@nevis.columbia.edu}

\begin{abstract}
Fully reconstructed jets and direct photon-tagged jet fragments significantly reduce energy-loss bias, the bias toward mostly measuring particles from partons which suffer little energy loss. In d+Au collisions, one accesses the physics at large $x$, which yields important constraints for nuclear parton distribution functions. In both d+Au and A+A collisions, coherent multiple-scattering models of energy loss can be tested. In this contribution, we present the current results from the PHENIX experiment on fully reconstructed jets and direct $\gamma$-hadron correlations. Baseline measurements of jets in p+p collisions as well as their yield and correlation modifications in d+Au and Cu+Cu will be given. From $\gamma$-hadron correlations, we present the fragmentation function in p+p and Au+Au collisions and its modification in Au+Au to $z_T$ lower than what has previously been studied. Implications of this data on our understanding of both cold and hot, dense nuclear matter created at RHIC are discussed.
\end{abstract}

\submitto{\jpg}
\maketitle

The PHENIX experiment has a mature program of measuring hard processes to perform precision QCD analysis and using these rare probes to characterize the medium produced at RHIC. In this contribution the current analysis of prompt $\gamma$-hadron correlations and fully reconstructed jets are discussed.

PHENIX has measured the azimuthal correlations between prompt photons and charged hadrons using two different methods. In p+p collisions, after removing photons event-by-event from hadronic decays, direct photons are measured using an isolation cut requiring that the sum of track $p_T$ and electromagnetic cluster energy in a cone of radius ($R=\sqrt{\Delta\phi^2+\Delta\eta^2}$) 0.3 be less than 10\% of the candidate photon's $p_T$\cite{ppgammajet}. Because of the large underlying event in Au+Au collisions, an isolation cut is difficult. Therefore, we have measured direct $\gamma$-hadron correlations statistically\cite{auaugammajet}. This is done by measuring the inclusive $\gamma$-hadron correlations and $\pi^0$-triggered correlations which are then mapped to decay photon correlations. Since we measure the ratio of inclusive to direct photons independently, we can subtract the decay-$\gamma$-triggered correlation from the inclusive to obtain the direct-$\gamma$-triggered correlations. Both methods yield similar results in p+p collisions giving evidence that the methods are robust.

\begin{figure}[b!]
\begin{center}
\includegraphics[width=0.45\textwidth]{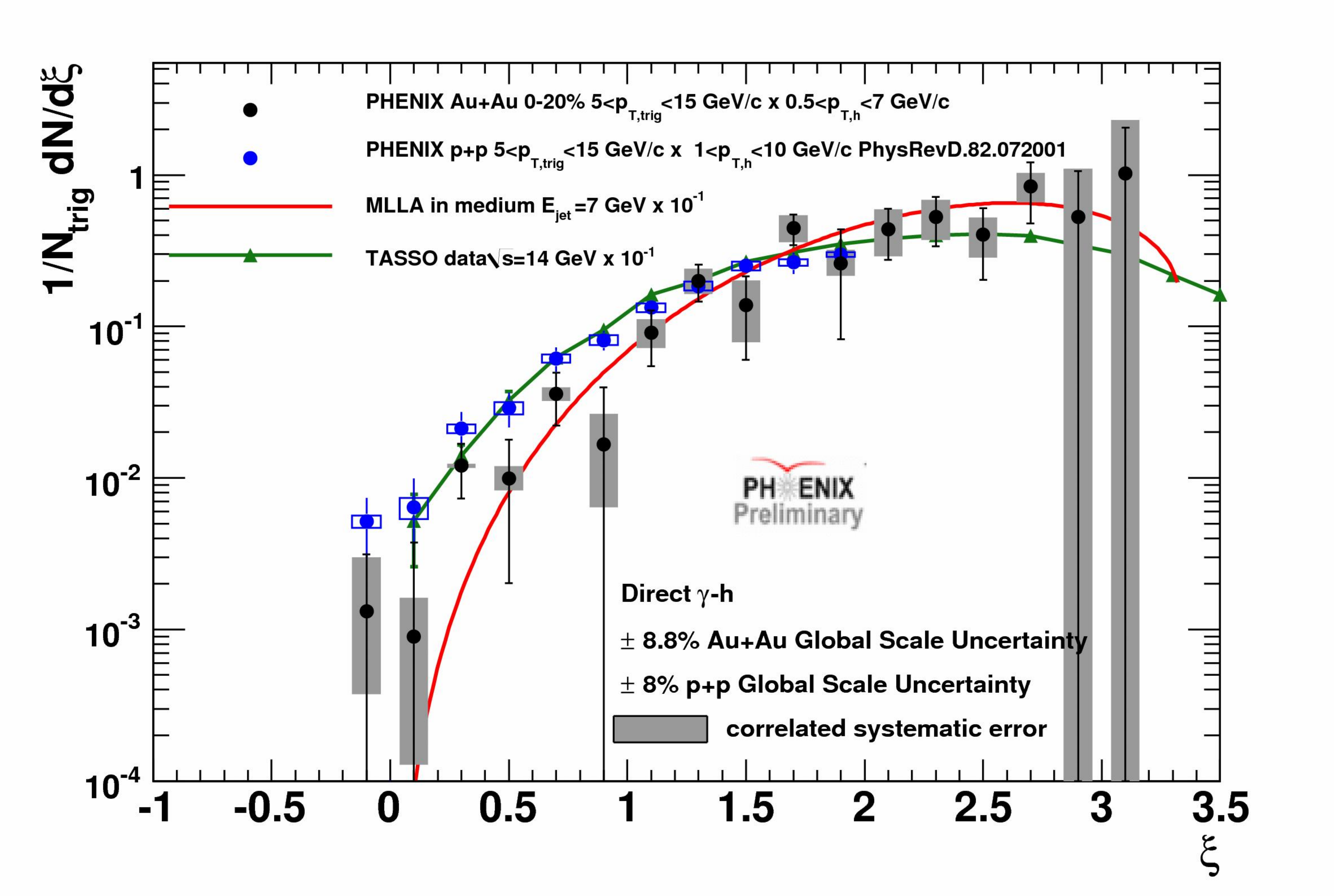}
\includegraphics[width=0.45\textwidth]{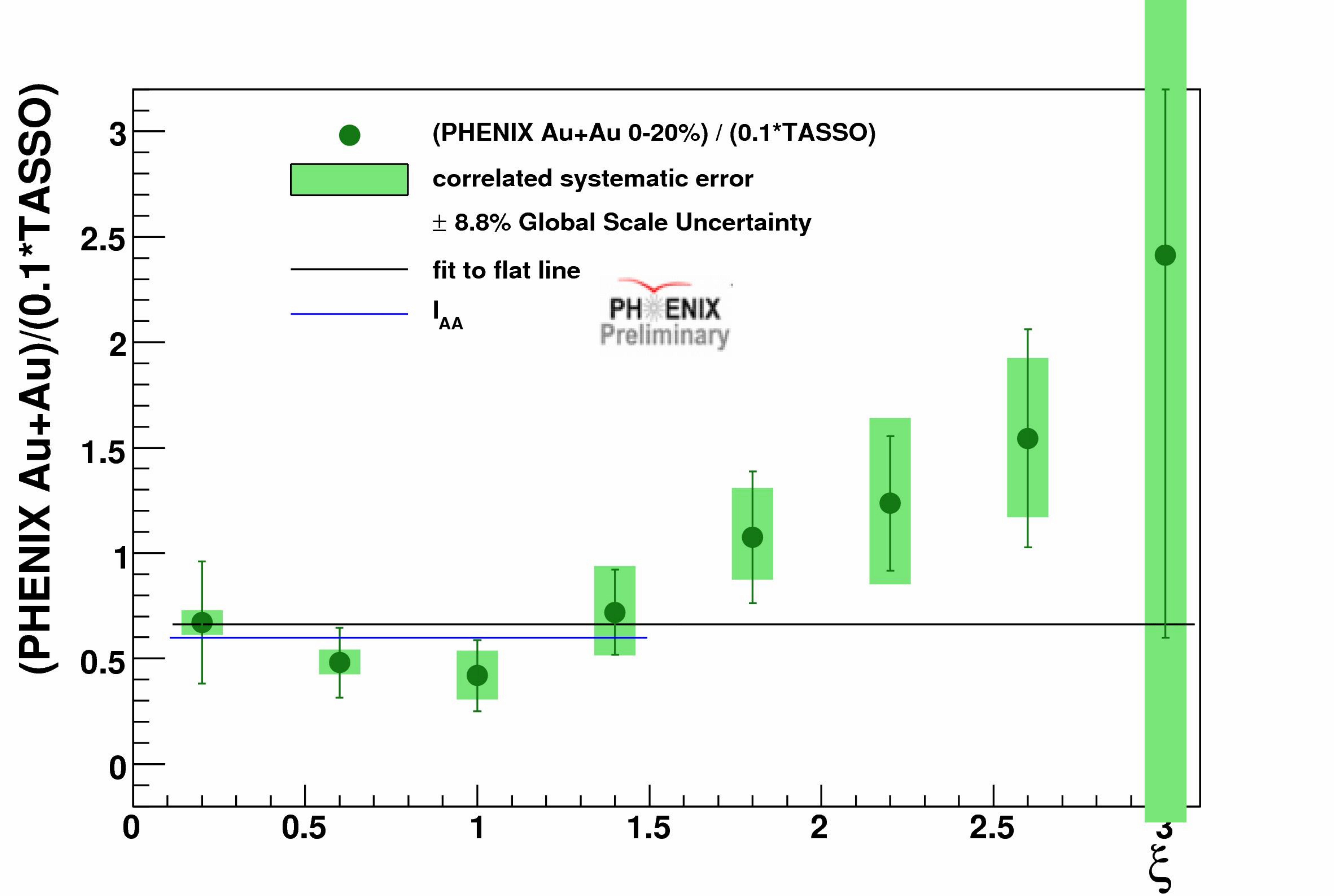}
\end{center}
\caption{(Left:)The yield of hadrons opposite to a direct photon in Au+Au (black circles) and p+p (blue circles). For comparison TASSO data (green line) and an energy-loss modified MLLA (red curve) is also shown. (Right:) The ratio of the Au+Au data divided by the TASSO data. The black line is a flat line fit to the data yielding a $\chi^2$/ndf = 12.16/7.} \label{fig:gammajetxi}
\end{figure}

Using these direct-photon-tagged correlations we can obtain the fragmentation function of the opposing jet, assuming the photon fixes the jet energy. We measure the per-trigger yield of hadrons correlated within $\Delta\phi>\pi/2$ from the direct photon as a function of 
\begin{equation}
\xi = \ln\left(\frac{p_{T}^{\mathrm{h}}\cos\Delta\phi}{p_{T}^{\gamma}}\right)
\end{equation}
This is shown for p+p and Au+Au collisions in Figure~\ref{fig:gammajetxi}. For comparison, data from TASSO $e^+e^-$ annihilation at $\sqrt{s} = $14 GeV is shown\cite{TASSO}. The p+p follow this TASSO data well lending credence to the fact that we are measuring a quark fragmentation function. This is interesting especially considering the addition of $k_T$ in p+p collisions relative to $e^+e^-$ annihilation which would modify the p+p with respect to $e^+e^-$. An arbitrarily normalized energy-loss modified MLLA fragmentation function \cite{MLLA} is plotted also and could be compared with the Au+Au data.

To see if there is any modification of the fragmentation function, we plot the ratio of the Au+Au fragmentation function to the TASSO fragmentation function. This is shown in the right panel of Figure~\ref{fig:gammajetxi}. We use this ratio because we have yet to measure the p+p fragmentation function down to a hadron $p_T$ of 0.5 GeV which yields $\xi$ values up to 3. What we see in this ratio is that a flat line fit to this ratio yields a poor $\chi^2/$ndf of 12.16/7 indicating that the shape of the fragmentation function is being modified in Au+Au collisions compared to TASSO data.

\begin{figure}[b!]
\begin{center}
\includegraphics[width=0.45\textwidth]{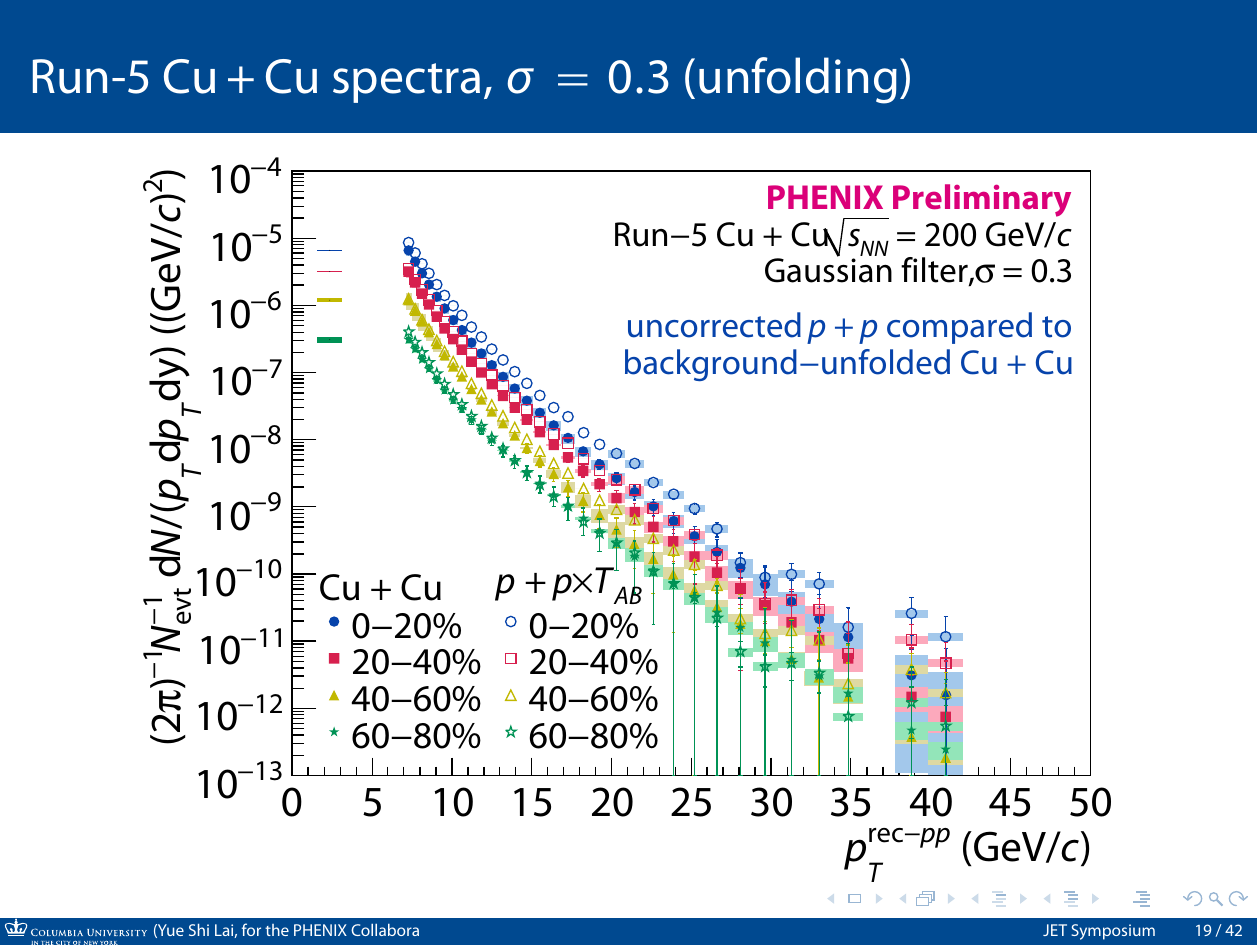}
\includegraphics[width=0.45\textwidth]{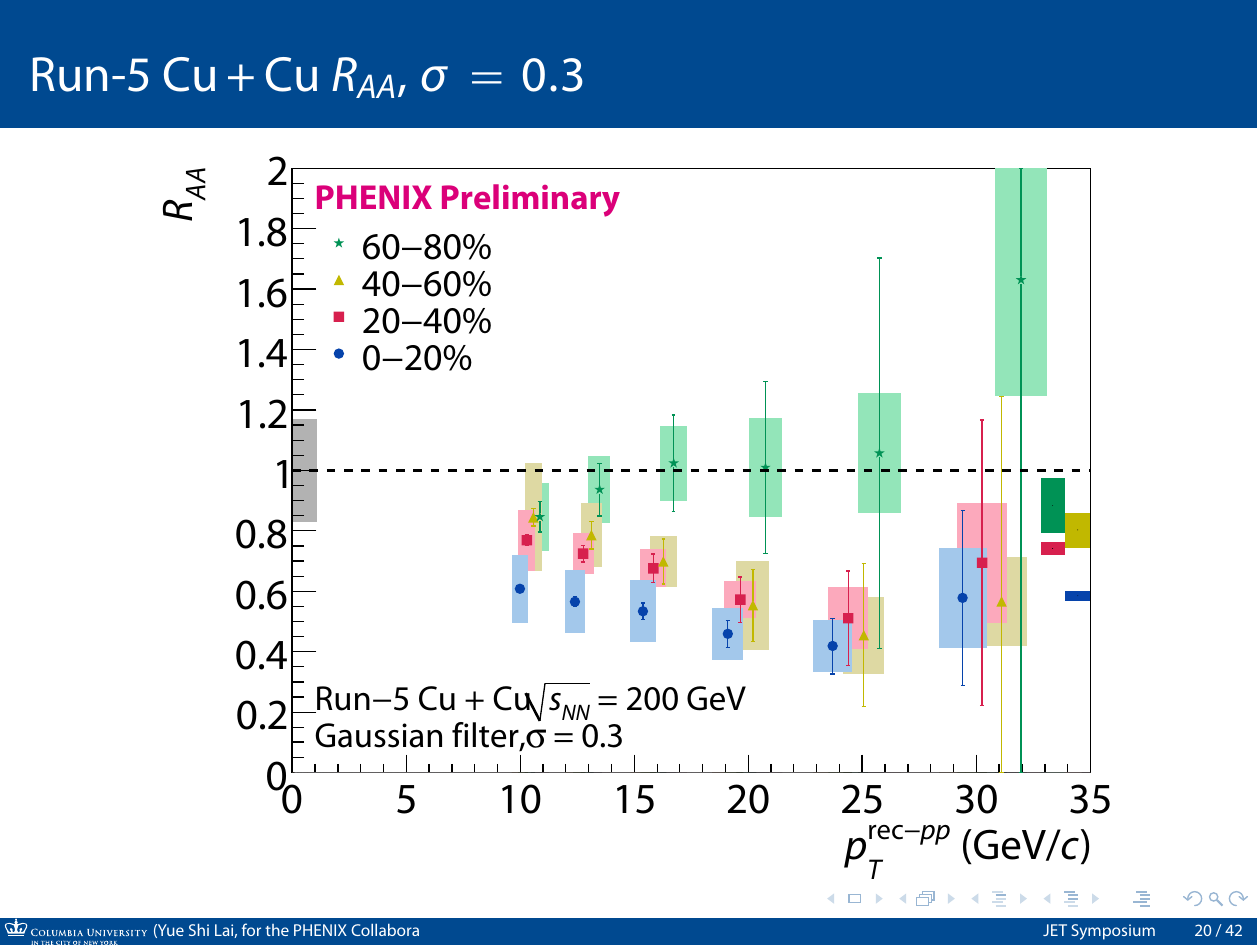}
\end{center}
\caption{(Left:) The invariant yield of Gaussian filter ($\sigma=0.3$ jets measured at the raw reconstructed p+p scale for $T_{AB}$-scaled p+p jets(open circles) and Cu+Cu (closed circles) for different centrality selections.}\label{fig:cucujets}
\end{figure}

Another way to fix the jet energy scale is to fully reconstruct jets. PHENIX has studied jet reconstruction using the Gaussian filter algorithm \cite{filterjet} in Cu+Cu and p+p collisions. When running jet reconstruction algorithms in a heavy ion environment, it is necessary to carefully take into account the effects of the underlying event. Beyond smearing the true energy and position of a jet,  more significantly, the underlying event particles can cluster themselves and produce a jet according to the algorithm's definition -- a so-called fake jet. In the Cu+Cu analysis, when the $p_T$ density of the particles are obtained, an event-averaged background, dependent on the centrality class, the z-vertex, and the reaction plane angle, is removed before the algorithm is run. After the algorithm is run, fake jets from the sample must be removed. This is done by a shape analysis of the jets where broad and diffuse jets are tagged and removed. The fake jet rejection cut is determined by evaluating its effectiveness at reducing the combinatorial contribution to di-jet $\Delta\phi$ distributions. To take into account the smearing of the underlying event, reconstructed jets from p+p collisions are embedded into minimum bias Cu+Cu events and the the algorithm is rerun. A transfer matrix is built based on the comparison of the input and output jets. This is used to unfold the effects of the underlying event so that the Cu+Cu is measured at the raw p+p reconstruction scale. The resulting jets at the p+p reconstruction scale are shown at the left panel of Figure.~\ref{fig:cucujets}.

Once the yields are obtained the nuclear modification factor, $R_{AA}$ is be measured. This is shown in the right panel of Figure~\ref{fig:cucujets}. There is a centrality-dependent suppression that is observed in the jet spectrum with the most peripheral events being consistent with no suppression. Although we present below new evidence of possible cold nuclear effects that may contribute partially to this observed suppression, other explanations are possible. For one, out-of-cone radiation results in a lower jet energy in Cu+Cu than p+p, reducing $R_{AA}$. Also, some of the removed fake jets are real, modified jets that are being excluded from the analysis. Either or both mechanisms like contribute to the measured suppression. But, a suppression in either case indicates modification of jets. We are currently exploring precisely how the jets are modified.

\begin{figure}
\begin{center}
\includegraphics[width=0.45\textwidth]{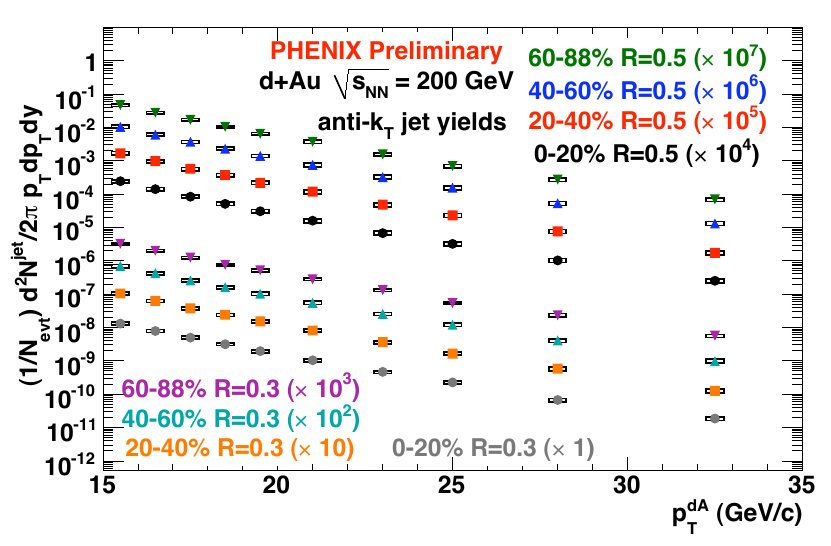}
\includegraphics[width=0.45\textwidth]{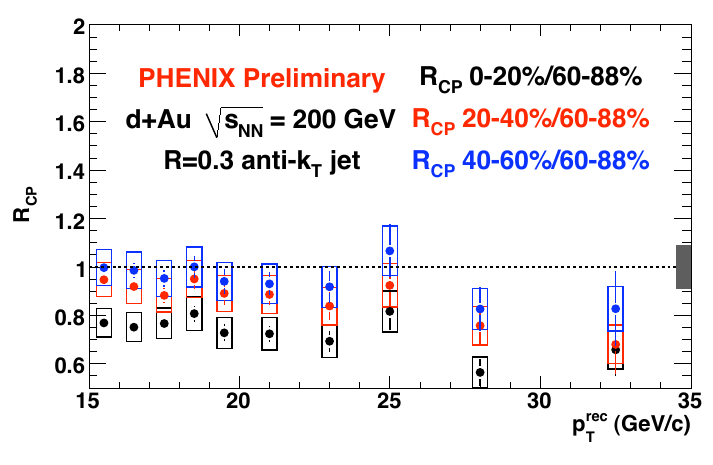}
\end{center}
\caption{(Left:) Yield of Anti-$k_T$ jets into the PHENIX acceptance for $R=0.3$ and $R=0.5$ jets for different centrality selections in d+Au. (Right:) The ratio of yields in central d+Au to those in peripheral d+Au.}\label{fig:daujets}
\end{figure}

We have also measured jets in d+Au collisions using the Anti-$k_T$ algorithm \cite{fastjet} to explore the effects of cold nuclear matter. The underlying event in d+Au collisions is less than Cu+Cu but not negligible. While subtracting an average underlying is not necessary, we are sensitive to correlated underlying event fluctuations that result in a jet according to the algorithm. Because of this we have studied jets with $R=0.3$ and $R=0.5$. To account for smearing from the underlying event, we use the same method as in Cu+Cu where p+p jets are embedded into minimum bias d+Au events and the jets are rerun and compared. The resulting spectrum of jets into the PHENIX acceptance at the raw reconstructed p+p scale is shown in the left panel of Figure.~\ref{fig:daujets}.

With the yields in hand, we can construct the ratio $R_{CP}$, the ratio of a central selection with the most peripheral selection. This is shown for $R=0.3$ jets on the right panel of Figure~\ref{fig:daujets}. What is seen is that the yield of jets is suppressed at the jet $p_T$ that we measure. This result is also consistent with the published single particle $\pi^0$ data\cite{pi0RdA}. Even though the $R_{dA}$ is consistent with unity for all centrality selections within the large systematic uncertainty, an $R_{CP}$ of the 0-20\% central to the 60-88\% peripheral selections do show a suppression of 20\%. This suppression could indicate either that some form of cold nuclear matter energy loss is suppressing jets at RHIC or that we are reaching $x$ values where the EMC effect is reducing the number of partons in the nuclear wavefunction\cite{eps09}. These considerations imply that we may need to re-interpret some of the high-$p_T$ suppression in Au+Au as being attributable to cold nuclear matter effects.


\end{document}